# Free Tuition, Stratified Pipelines: Four Decades of Administrative Cohorts and Equity in Access to Engineering and Science in an Argentine Public University

Hugo Roger Paz
PhD Professor and Researcher Faculty of Exact Sciences and Technology National University of Tucumán
Email: hpaz@herrera.unt.edu.ar
ORCID: https://orcid.org/0000-0003-1237-7983

**Abstract**

Latin American higher education is often portrayed as equitable when tuition is free and access to public universities is formally unrestricted. Yet, a growing body of research shows that massification under **tuition-free policies** can coexist with strong social and territorial stratification. This article uses four decades of administrative records from a faculty of engineering and exact sciences in north-western Argentina to examine who manages to enter these programmes, and how cohort composition has changed over time.

Drawing on 24,133 first-time entrants (1980–2019), we construct a leakage-aware "background census" layer (N1c) that harmonises secondary school type and management, province of origin, and age at entry across successive generations of legacy systems. We then combine descriptive analyses by decade, UMAP+DBSCAN clustering of background profiles, and a reconstructed macroeconomic panel (inflation, unemployment, poverty, GDP) anchored in the year of entry. All analyses are descriptive and explicitly report structural missingness patterns.

Results show, first, that missingness in school and geographic variables is historically patterned rather than random, declining sharply after the 1990s. Second, among students with observed school information, the share coming from private—especially religious and bilingual—secondary schools located in high-income areas such as Yerba Buena increases from less than half in the 1980s to around two-thirds in the 2010s. Third, the faculty's **catchment area** becomes more local, with the home province gaining weight among entrants, while neighbouring provinces and distant origins lose ground. Median age at entry remains stable at 19 years, with persistent right tails of older entrants. Macro-linkage analyses reveal moderate associations between unemployment and older age at entry, and between inflation and a higher share of students from interior provinces.

We argue that, in this setting, free tuition and open entry have operated within, rather than against, stratified school and residential pipelines. The article illustrates how leakage-aware use of administrative data can support equity monitoring at the faculty level and discusses implications for upstream school policies, macro-sensitive support instruments, and institutional accountability in **tuition-free systems**.

**Keywords**

higher education equity, access to engineering and science, administrative data, social stratification, Latin America / Argentina, free tuition and public universities.

# 1. INTRODUCTION

## 1.1. Higher education massification under persistent inequality

Over the last four decades, higher education in Latin America has expanded rapidly, with enrolment growing in both public and private institutions and participation rising among groups historically excluded from university spaces (Arias Ortiz et al., 2020; Villalobos et al., 2017). Yet this massification has unfolded in deeply stratified systems, where strong segmentation between public and private provision and pronounced school segregation by socio-economic status reproduce inequality in access to prestigious programmes, including engineering and the physical sciences (Gasparini et al., 2011; Murillo & Martínez-Garrido, 2019; Villalobos et al., 2017).

Empirical work using household surveys and international assessments shows that Latin America is among the regions with the highest levels of school segregation, and that the expansion of private schooling has often increased, rather than reduced, socio-economic stratification (Gasparini et al., 2011; Murillo & Martínez-Garrido, 2019). At the tertiary level, access to selective public universities and high-return programmes remains strongly associated with social origin, school sector and place of residence (Arias Ortiz et al., 2020; Villalobos et al., 2017). In this context, describing *who* manages to enter engineering and science programmes is central to understanding whether massification has translated into greater equity or into new forms of stratified access.

Latin American higher education has also evolved under recurrent macroeconomic crises, labour-market volatility and episodes of sharp inflation. Evidence from Argentina, Brazil and Mexico indicates that negative income shocks significantly increase the risk that young people exit upper secondary and tertiary education (Cerutti et al., 2019). Regional reviews highlight that financial constraints and unstable employment remain key barriers to completion, particularly for low-income students (Josephson, 2018). These conditions mean that university access is shaped not only by long-term social stratification but also by short-term macroeconomic turbulence.

## 1.2. Administrative data and the challenge of understanding trajectories

Universities increasingly rely on administrative records to monitor dropout, design early-warning systems and evaluate interventions. Recent work using newly established data systems in Guatemala and Honduras, for example, has shown that administrative data can support predictive models of school dropout and inform targeting strategies in low- and middle-income countries (Adelman et al., 2018). Similar initiatives in Latin America and beyond underline the potential of longitudinal records and learning analytics for identifying high-risk trajectories and supporting timely policy responses (Heredia-Jiménez et al., 2020; IDRC–ILDA, 2024).

However, most institutional analytics remain constrained by short observation windows and opportunistic feature sets shaped by what happens to be stored in legacy systems rather than by an explicit theoretical design (Adelman et al., 2018; Paz, 2025). In many universities, background information such as school of origin, school management, province of residence or age at entry is recorded unevenly across programmes and decades, with high levels of missingness and shifting coding conventions. Left untreated, this "administrative chaos" hampers longitudinal analyses of student trajectories and biases any attempt to describe who gains access to specific programmes.

Within the CAPIRE framework, recent methodological work has proposed a leakage-aware data layer for student analytics that explicitly separates observation windows from outcome horizons and organises predictors into multilevel feature sets (N1–N4) (Paz, 2025). This approach shows how careful data engineering—combined with concepts such as the Value of Observation Time—can turn fragmented administrative archives into coherent trajectory cohorts. Building on that foundation, the present manuscript shifts the focus from predictive performance to descriptive insight: given a cleaned and leakage-aware background layer, *what can we learn about who our students are when they cross the institutional threshold?*

### 1.3. Why entering cohorts matter

Entry cohorts condense decades of social selection. By the time a student enrols in an engineering or science programme at a public university, multiple filters have already operated: family socio-economic position, school quality, geographic distance to the institution, gender norms and the opportunity cost of continued study (Gasparini et al., 2011; Villalobos et al., 2017). Cohort composition therefore provides a sensitive indicator of whether higher education is functioning as a mechanism for social mobility or for the reproduction of existing hierarchies.

Three dimensions are particularly salient in the Latin American context. First, **school sector and school type** matter because private and public schools, as well as religious and secular institutions, often cater to very different social groups and track students into stratified educational and occupational pathways (Gasparini et al., 2011; Murillo & Martínez-Garrido, 2019). Second, **geographic origin** interacts with labour markets and institutional geographies: the distribution of public and private universities across territories shapes who can realistically access certain programmes, and spatial mismatch between local labour markets and higher education supply has been documented in several Latin American countries (Arias Ortiz et al., 2020). Third, **socio-economic proxies and age at entry** provide indirect but crucial information, since direct measures of family income are rarely available in administrative data. Age at entry, in particular, can signal interrupted school trajectories, previous attempts at higher education or the need to combine study with work—patterns that become more prevalent when households face income shocks (Cerutti et al., 2019).

**1.4. From snapshots to longitudinal cohorts in a volatile economy**

Most descriptions of student background in Latin America rely on cross-sectional surveys or short administrative snapshots (Arias Ortiz et al., 2020; Villalobos et al., 2017). While these approaches are valuable, they have two limitations for our purposes. First, they seldom follow a single institution over multiple decades with consistent definitions, making it difficult to disentangle institutional dynamics from national or regional trends. Second, they rarely integrate macroeconomic context—such as inflation, unemployment or poverty dynamics—into cohort descriptions, even though a growing literature shows that economic crises alter schooling decisions, particularly for young people at the margin of higher education (Cerutti et al., 2019; Josephson, 2018).

The present study addresses both limitations. We use a leakage-aware longitudinal dataset of all first-time entrants to the Faculty of Exact Sciences and Technology (FACET) at the National University of Tucumán from 1980 to 2019, harmonising key background variables—school type and management, province of origin and age at entry—across deep historical changes in coding practices (Paz, 2025). In parallel, we assemble a macroeconomic panel for Argentina (1980–2019) with indicators of inflation, unemployment, poverty and real GDP per capita, based on official statistics and international databases. This allows us to explore whether shifts in cohort composition—for example, the share of entrants from interior provinces or from private schools—coincide with periods of acute macroeconomic turbulence.

**1.5. Aim and contributions of this manuscript**

This manuscript is the first in the CAPIRE 2 series to focus explicitly on **who** enters a Latin American faculty of engineering and exact sciences, and **how that composition has changed over four decades**. Using the cleaned N1c background layer and a set of seven complementary analyses, we pursue three overarching aims.

First, **to reconstruct the evolution of recorded student background under severe data constraints**. We quantify and model historical missingness in key fields such as school type, showing that it is structurally embedded in earlier decades and specific programmes rather than random noise.

Second, **to describe long-term patterns in school sector, geographic origin, socio-economic proxies and age at entry**. We characterise how the balance between public and private schools, local versus interior provinces, and younger versus older entrants has shifted across enrolment decades, and whether there is any stable "typical" profile of the engineering student.

Third, **to locate these cohort changes within macroeconomic and institutional context**. By linking the N1c layer to a reconstructed macro panel for 1980–2019, we provide initial evidence on how inflation and labour-market conditions correlate with composition indicators such as the proportion of students from the interior or the average age at entry, without making causal claims.

Substantively, the study offers a rare four-decade institutional portrait of entering cohorts in a public faculty of engineering and exact sciences in the Argentine Northwest, situated within broader debates on equity and the right to higher education in Latin America (Josephson, 2018; Villalobos et al., 2017). Methodologically, it illustrates how a leakage-aware trajectory framework can turn messy administrative archives into analytically coherent cohorts that are comparable over time and linkable to macro context (Adelman et al., 2018; Paz, 2025). The remainder of the paper details the institutional setting and data sources (Section 2), the analytical strategy (Section 3), the seven background analyses (Section 4) and their implications for equity-oriented policy and future CAPIRE-based modelling (Sections 5–7).

## 2. INSTITUTIONAL AND DATA CONTEXT

### 2.1. Institutional setting

This study draws on administrative records from a public engineering-oriented faculty embedded in a large Argentine national university system characterised by tuition-free access, strong regional roots and a long tradition of training engineers and scientists. In line with broader Latin American trends, the institution operates within a massified yet highly stratified higher-education landscape, where public universities coexist with a rapidly expanded private sector and increasing institutional differentiation (Brunner & Miranda, 2016; Ferreyra et al., 2017; Levy, 1986).

Within this ecosystem, the faculty of exact sciences and engineering hosts multiple undergraduate degree programmes in engineering, basic sciences and applied technologies. Programmes typically follow long, credit-intensive curricula with strong mathematics and physics foundations, mirroring international patterns in engineering education where rigid structures and high early attrition are the norm (Graham, 2018). The institution admits students primarily from the province and surrounding regions in north-western Argentina, but also receives entrants from other parts of the country, especially in engineering programmes with historical prestige.

From an information-systems perspective, the faculty's student data infrastructure is the result of successive generations of administrative software that were initially designed for operational tasks (enrolments, grades, certificates) rather than research or analytics. As in many Latin-American universities, this has produced heterogeneous coding schemes, parallel legacy subsystems and uneven standardisation of key descriptors such as school type, geographic origin or socio-economic variables, with gradual improvements in data quality over the last four decades (Brunner, 1991; INDEC, 2019).

**2.2. N1c census-like background layer and cohort definition**

The empirical core of this manuscript is the N1c "background census" layer of the CAPIRE framework (Paz, 2025). This layer condenses, for each first-time entrant, the information that is available at (or before) the moment of initial enrolment in any undergraduate programme of the faculty between 1980 and 2019. Conceptually, N1c sits immediately above the raw person-level and censal tables of the institutional database, and below curriculum-level and event-level analytic layers described elsewhere (Paz, 2025).

The construction of N1c proceeds in three steps:

1. **Person consolidation.** Records from legacy systems are first merged into a students_master file that resolves duplicates, harmonises identifiers and links each individual to one or more degree programmes and administrative trajectories. This step is necessary because, over four decades, students may have multiple IDs, change programmes or re-enter after interruptions.

2. **Selection of analytic universe.** From students_master, we extract a cohort file containing one record per individual linked to their *first* degree programme in the faculty, restricted to initial enrolments between 1980 and 2019. This produces a longitudinal universe of approximately twenty-four thousand entrants, covering the transition from elite to massified access in Argentine engineering and related fields.

3. **Projection to N1c.** The N1c layer retains only variables that are: (a) known at or before first enrolment (e.g., sex, age, secondary school, declared place of residence), and (b) sufficiently populated to support longitudinal analyses after careful normalisation. All identifiers are pseudonymised before export, and no directly identifying information (names, document numbers, precise addresses) is included.

A crucial design principle is temporal "non-leakage": N1c does not incorporate any information that depends on post-entry events (subject approvals, exam failures, subsequent programme changes, etc.). This ensures that the background analyses in this manuscript reflect only ex-ante characteristics, preserving the validity of later predictive and causal models that will use N1c as a baseline (Paz, 2025).

**2.3. Background variables and normalisation**

The N1c layer includes three broad families of variables relevant for this paper: cohort timing, pre-university educational background and geographic origin, complemented by basic demographic attributes.

**Cohort timing.**

Each student is assigned an entry year and, for descriptive purposes, to an entry decade (1980s, 1990s, 2000s, 2010s). Cohort decade is used to summarise long-term patterns in school composition, geographic origin, socio-economic proxies and age at entry.

**Educational background.**

Secondary schooling is captured through a combination of free-text descriptors and coded fields in the original administrative systems (school name, location, type and managing authority). These are normalised into canonical categories that distinguish:

- **School orientation** (e.g., technical vs non-technical/academic),
- **School management** (public vs private), and
- **Administrative dependency** (national vs provincial, where recoverable).

Because school descriptors evolved substantially over time, the normalisation pipeline uses a combination of dictionary-based harmonisation, fuzzy matching and manual verification for high-frequency institutions, followed by aggregation into robust categories that can be tracked across decades. The result is a set of indicators that describe, for each cohort and decade, the relative weight of technical versus non-technical schools and public versus private management among entrants. At the individual level, these variables later feed composite proxies of socio-economic position used in the analyses in Section 4.

From a **contextual socio-economic perspective**, the public–private distinction is not interpreted abstractly. In the faculty's catchment area, private secondary schools include a small group of **religious and bilingual institutions located in high-income zones such as Yerba Buena** (for example, long-established colegios with international or bilingual curricula), which typically cater to upper-middle-class and affluent households. Public schools, by contrast, enrol a much broader range of social backgrounds and, in aggregate, represent predominantly middle- and lower-middle-income students. When we use private schooling as a proxy for higher socio-economic status in later sections, it should therefore be read as a

**conservative indicator of belonging to these upper-middle-class circuits**, rather than as a literal measure of income.

**Geographic origin.**

Geographic background is derived from the province (and, where available, department) of the secondary school and from declared place of residence at entry. Given substantial inconsistencies in early years, the normalisation focuses on achieving a stable province-level classification, with particular attention to:

- the home province of the institution,
- surrounding provinces in the north-west region, and
- other provinces and jurisdictions (including the Buenos Aires metropolitan area).

These variables allow us to quantify the long-term evolution of the faculty's catchment area and the shifting balance between local and extra-provincial students, which is central for interpreting equity and access in a massified system.

**Demographic attributes.**

Finally, N1c includes sex (binary coding as recorded in the administrative systems) and age at entry, calculated as the difference between date of birth and date of first enrolment. Age at entry is treated as a continuous variable and summarised by decade and school type, providing an additional window into how macroeconomic pressures and changing school pathways shape the timing of university access.

.2.4. Macroeconomic panel and linkage strategy

To situate these background patterns in their broader socio-economic environment, we assemble a companion macro panel for Argentina covering 1980–2019 (macro_1980_2019.csv). The panel includes annual series for:

- inflation (consumer prices, annual %),
- unemployment rate (total, % of labour force),
- poverty headcount (% of population, national lines where available), and
- real GDP or real GDP per capita (constant-price series or annual growth rates).

The series are reconstructed from official and international statistical sources, combining World Bank World Development Indicators, SEDLAC/CEDLAS–World Bank distributional statistics and Argentina's own historical statistical releases (INDEC). For each indicator we document, in a separate metadata file, the precise source, units, coverage, known methodological breaks and any splicing decisions undertaken to obtain a consistent 1980–2019 trajectory.

The linkage between the macro panel and N1c is deliberately simple and transparent: each entry cohort is associated with the calendar-year value of the macro indicators corresponding to its year of first enrolment. This "year-of-entry" anchoring reflects the economic environment that families and prospective students face when deciding whether to begin an engineering-related degree, rather than subsequent shocks during the trajectory. In Section 5, we exploit this linkage to explore how long-term swings in inflation, unemployment and poverty covary with shifts in the composition of entrants by school type, geographic origin and age at entry.

By integrating a carefully normalised census-like background layer (N1c) with a documented national macro panel, the data architecture for this manuscript provides a coherent bridge between individual-level educational histories and the broader economic tides that shape who arrives at the gates of engineering and science programmes in a massified yet unequal system.

## 3. METHODS

### 3.1. Overall analytical strategy

Our analytical strategy is descriptive, longitudinal and explicitly leakage-aware. Building on the N1c background layer described in Section 2 and the CAPIRE data architecture (Paz, 2025), we structure the work into seven complementary analyses that jointly characterise four decades of entering cohorts:

1. **Structural missingness in school-type variables** (Analysis 1).
2. **School-type composition by decade** (Analyses 2–3).
3. **Geographic origin and catchment evolution** (Analyses 4–5).
4. **Socio-economic proxies and age at entry** (Analyses 6–7).
5. **Low-dimensional cohort structure and macro linkages** (Analyses 6–7, extended).

Analyses are conducted at two levels:

- **Cohort–decade summaries**: proportions by school type, school management, province groups and deciles of age at entry.

- **Individual-level structure**: cluster analysis in a reduced UMAP space, treating decades, school-type categories and missingness indicators as joint descriptors.

All computations were implemented in Python, using pandas and numpy for data wrangling, and scikit-learn and umap-learn for modelling and clustering (McInnes et al., 2018; Pedregosa et al., 2011).

### 3.2. Handling missing data and structural missingness

#### 3.2.1. Conceptual stance

We treat missing data not as a nuisance to be silently "fixed", but as a signal of historical and institutional processes. Following the standard taxonomy, we distinguish between data missing completely at random (MCAR), missing at random (MAR) and missing not at random (MNAR), and emphasise that naive complete-case analyses can be severely biased when these conditions fail (Little & Rubin, 2019).

In the N1c layer, missingness is concentrated in key variables describing secondary school type and management, especially in the 1980s and early 1990s. Our aim is therefore twofold:

1. **To model missingness itself** as an outcome explained by observable features (entry decade, programme family, sex, province of origin).

2. **To avoid imputation for core descriptive results**, using observed data only, but making explicit which decades and variables are affected by structural gaps.

#### 3.2.2. Operationalisation

For each background variable $X$ (e.g., technical vs. non-technical school), we define a binary missingness indicator $M_X$ that equals 1 when the field is missing or unusable, and 0 otherwise. We then estimate logistic regression models of the form:

$$\Pr(M_X = 1 \mid Z) = \text{logit}^{-1}(\alpha + \beta^\top Z),$$

where $Z$ includes:

- decade of entry (categorical),
- degree family (engineering vs. other programmes),
- sex, and
- province group (local vs. interior vs. other provinces).

These models are estimated using maximum likelihood with standard errors clustered by entry year to account for within-year correlation. Goodness of fit is assessed via pseudo-$R^2$, area under the ROC curve and calibration plots (Hastie et al., 2009).

We interpret high predictive performance of $M_X$ as evidence of **structural missingness** driven by observable factors, rather than idiosyncratic data entry failures. This is the basis for the claim, used later in the discussion, that early-decade missingness in school-type variables is historically structured rather than random noise.

No multiple imputation is used for the main descriptive analyses. Instead, we:

- report **denominators** explicitly for each percentage (e.g., number of entrants with observed school type in a given decade),
- flag decades where coverage falls below pre-defined thresholds, and
- restrict some analyses (e.g., cluster modelling) to subperiods with sufficiently complete data.

This conservative stance avoids injecting modelled values into the cohort descriptions and preserves N1c for later work where imputation, if used, will be documented separately.

**3.3. Descriptive analyses by decade**

Analyses 1–5 rely on cohort–decade summaries:

- **Analysis 1 (missingness evolution).** We compute, for each entry year and decade, the proportion of records with missing school-type or management information, stratified by degree family and sex. This produces the time series underlying the missingness figure and the decade-level table of coverage.
- **Analyses 2–3 (school sector and type).** Among records with observed values, we estimate the share of entrants from:
    - public vs. private schools,
    - technical vs. non-technical schools, and
    - combinations of sector and orientation (e.g., public–technical, private–non-technical).

These are summarised by entry decade, with 95% confidence intervals for proportions computed using the Wilson method.

- **Analyses 4–5 (geographic origin).** We classify students according to:
    - local province (where the faculty is located),
    - interior provinces of the same macro-region, and
    - other provinces and jurisdictions (including the Buenos Aires metropolitan area).

We then compute decade-level shares and change indices relative to the 1980s baseline.

- **Analysis 6 (socio-economic proxies and age).** We construct simple proxies by combining school management (public/private), school orientation (technical/non-technical) and province group into categorical "background profiles". In parallel, we summarise age at entry via medians, interquartile ranges and kernel density estimates by decade and school-type categories.

All descriptive statistics are computed with standard pandas routines, and visualisations use standard plotting backends; no weighting is applied, as N1c is a census-like layer for the faculty.

### 3.4. UMAP-based clustering of background profiles

To explore whether entering cohorts exhibit recurrent configurations beyond simple marginals, we perform a low-dimensional embedding followed by density-based clustering (Analysis 7, structural).

### 3.4.1. Input representation

We restrict the clustering sample to entry years and variables with high coverage (post-1990s core). Each student is represented by:

- decade of entry (one-hot encoded),
- school orientation (technical vs. non-technical),
- school management (public vs. private),
- province group, and
- indicators for key missingness patterns (e.g., whether school management is missing).

Categorical variables are one-hot encoded into a sparse binary feature matrix.

### 3.4.2. UMAP embedding

We apply Uniform Manifold Approximation and Projection (UMAP) for nonlinear dimension reduction to a two-dimensional latent space (McInnes et al., 2018). Default parameters are tuned to balance local and global structure:

- n_neighbors = 15,
- min_dist = 0.1,
- metric = "euclidean".

UMAP is chosen over alternatives such as t-SNE because of its better preservation of global structure and suitability for larger samples while still producing interpretable 2D layouts.

### 3.4.3. Density-based clustering

On the 2D UMAP embedding, we run DBSCAN (Density-Based Spatial Clustering of Applications with Noise), a clustering algorithm that identifies high-density regions of arbitrary shape and labels points in low-density areas as noise (Ester et al., 1996).

We select DBSCAN hyperparameters using a **k-distance graph** approach:

1. For each point, we compute the distance to its $k$-th nearest neighbour in the UMAP space (with $k$ aligned with the min_samples parameter).
2. We sort these distances and inspect the curve to identify an "elbow" point that represents a transition between dense clusters and sparse regions.
3. We set eps at the elbow distance and min_samples in line with typical practice and sample size.

This procedure supports a transparent choice of eps instead of arbitrary trial-and-error, consistent with the original DBSCAN paper.

Cluster membership is then cross-tabulated with entry decade, school sector, school orientation and province group, providing the basis for the cluster-composition table and the UMAP visualisations.

### 3.5. Macro–N1c linkage and correlation analysis

The final analytical step links the N1c cohort summaries to the macroeconomic panel described in Section 2.4. For each entry year (1980–2019) we compute:

- the proportion of entrants from interior provinces,
- the proportion from private secondary schools,
- the share of technical-school graduates, and

- the mean age at entry.

These are merged with year-specific macro indicators:

- inflation rate,
- unemployment rate,
- poverty headcount, and
- real GDP per capita.

To characterise associations between macro conditions at entry and cohort composition, we compute:

- **Pearson correlation coefficients** for approximately linear relationships between macro indicators and continuous cohort measures (e.g., average age at entry), and
- **Spearman rank correlations** when monotonic but potentially non-linear relationships are plausible (e.g., inflation versus share of interior students), following standard definitions (Hastie et al., 2009; Spearman, 1904).

For each pair of variables we report:

- correlation coefficient (Pearson $r$ or Spearman $\rho$),
- two-sided p-value under the null of zero association, and
- 95% confidence intervals obtained from standard large-sample approximations.

Given the modest number of years, we interpret these correlations as **descriptive covariations**, not as causal effects. All analyses use standard functions from scipy/pandas and scikit-learn (Pedregosa et al., 2011). The resulting statistics feed directly into the macro–N1c correlation table and the multi-panel figure of trends and associations presented in Section 5.

## 4. RESULTS

### 4.1. Structural missingness in school-type data (Analysis 1)

The N1c layer comprises **24,133 first-time entrants** between 1980 and 2019 (plus a single record from the 1970s kept for completeness). However, the availability of secondary school information is highly uneven across decades. As shown in Figure 1, the **share of missing school-type records** is almost total in the early period and declines sharply over time:

- 1970s: 100% missing (1/1).
- 1980s: 98.1% missing (3,854/3,927).
- 1990s: 78.0% missing (3,621/4,644).
- 2000s: 32.4% missing (2,989/9,215).
- 2010s: 0.06% missing (4/6,346).

**Figure 1. Evolution of structural missingness in secondary school type records by entry decade (1970–2019).**

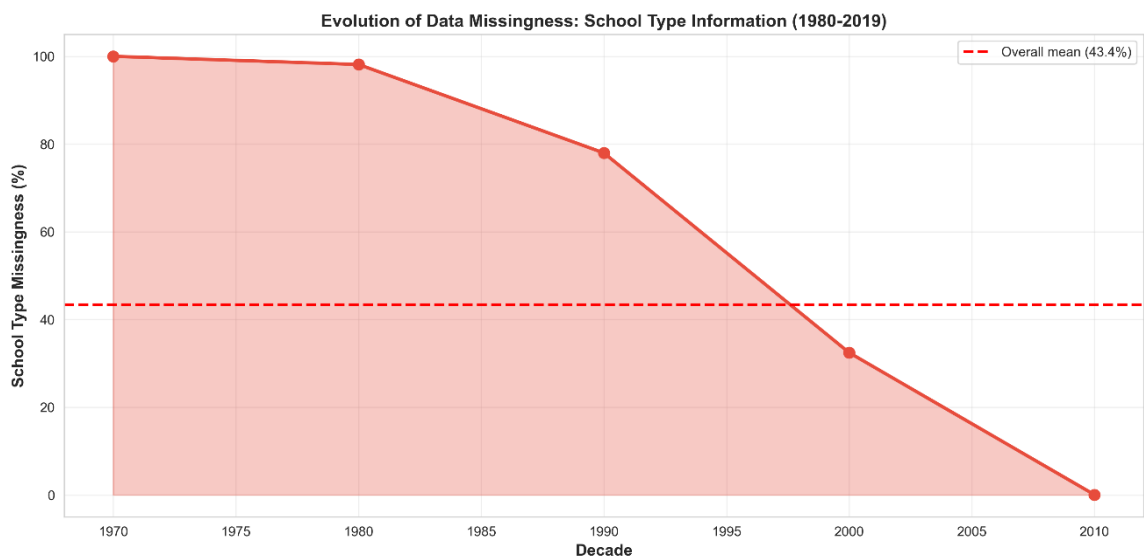

**Table 1. Magnitude and evolution of missing data in school type records by decade.**

| decade | missing_count | total_count | missing_rate | missing_pct |
|---|---|---|---|---|
| 1970 | 1 | 1 | 1.0 | 100.0 |
| 1980 | 3854 | 3927 | 0.9814 | 98.14 |
| 1990 | 3621 | 4644 | 0.7797 | 77.97 |
| 2000 | 2989 | 9215 | 0.3244 | 32.44 |
| 2010 | 4 | 6346 | 0.0006 | 0.06 |

Logistic models of missingness (not shown in detail here) indicate that whether school-type information is missing can be predicted almost perfectly (≈99.7% accuracy) from **entry decade, degree family and province coding**, confirming that missingness is **structural and historically patterned**, not random noise. The same temporal pattern appears in the province-of-origin field: in the 1980s, 98% of entries have no usable province, falling to 78% in the 1990s, 33% in the 2000s and virtually zero in the 2010s. Together, these results justify treating early-decade missingness as a legacy of administrative practices rather than as individual non-response, and motivate the focus on **decade-level** summaries and explicit denominators in subsequent analyses.

**4.2. School type and management (Analyses 2–3)**

**4.2.1. Composition by detailed school type**

Among entrants with observed school information, the distribution of **school type** changes markedly across decades (Table 2, Figure 2). In the 1980s, almost half of the documented entrants come from **state national schools** (49.3%), with smaller shares from **private religious** (19.2%), **private secular** (27.4%) and **state provincial** schools (4.1%). By the 2010s, this pattern is reversed:

- **Private religious**: from 19.2% (1980s) to 36.3% (2010s).
- **Private secular**: from 27.4% to 29.1% (relatively stable).
- **State national**: from 49.3% down to 23.8%.
- **State provincial**: increasing from 4.1% to mid-teens in the 1990s–2000s, then stabilising around 10.7% in the 2010s.

**Figure 2. Distribution of entrants by detailed secondary school type across decades (observed cases only).**

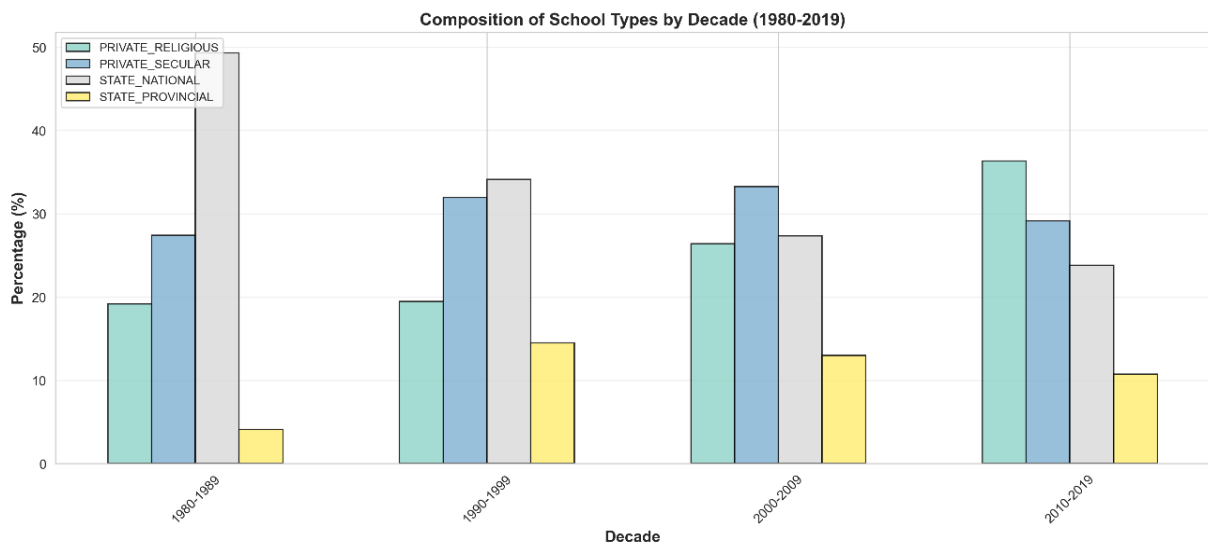

**Table 2. Relative frequency of secondary school types among entrants with observed data (1980–2019).**

| Decade | PRIVATE_RELIGIOUS | PRIVATE_SECULAR | STATE_NATIONAL | STATE_PROVINCIAL |
|---|---|---|---|---|
| 1980 | 19.18 | 27.40 | 49.32 | 4.11 |
| 1990 | 19.45 | 31.96 | 34.12 | 14.47 |
| 2000 | 26.41 | 33.28 | 27.34 | 12.98 |
| 2010 | 36.33 | 29.14 | 23.81 | 10.72 |

Thus, within the documented regime, **private religious schools** become the single largest category of origin by the 2010s, while **national state schools** lose more than half their relative weight. The evolution is gradual rather than abrupt, pointing to a long-run reconfiguration of the secondary school pipeline feeding engineering and science programmes.

### 4.2.2. Public vs. private management

Collapsing school types into **public vs. private management** reveals a clear shift towards private schooling among entrants (Table 3, Figure 3). In the 1980s, the documented cohort is slightly dominated by public schools (53.4% public vs. 46.6% private). By the 2010s, the pattern has flipped:

- 1980s: 46.6% private, 53.4% public.
- 1990s: 51.4% private, 48.6% public.
- 2000s: 59.7% private, 40.3% public.
- 2010s: 65.5% private, 34.5% public.

**Figure 3. Long-term shift in school management origin: Public vs. Private secondary schools (1980–2019).**

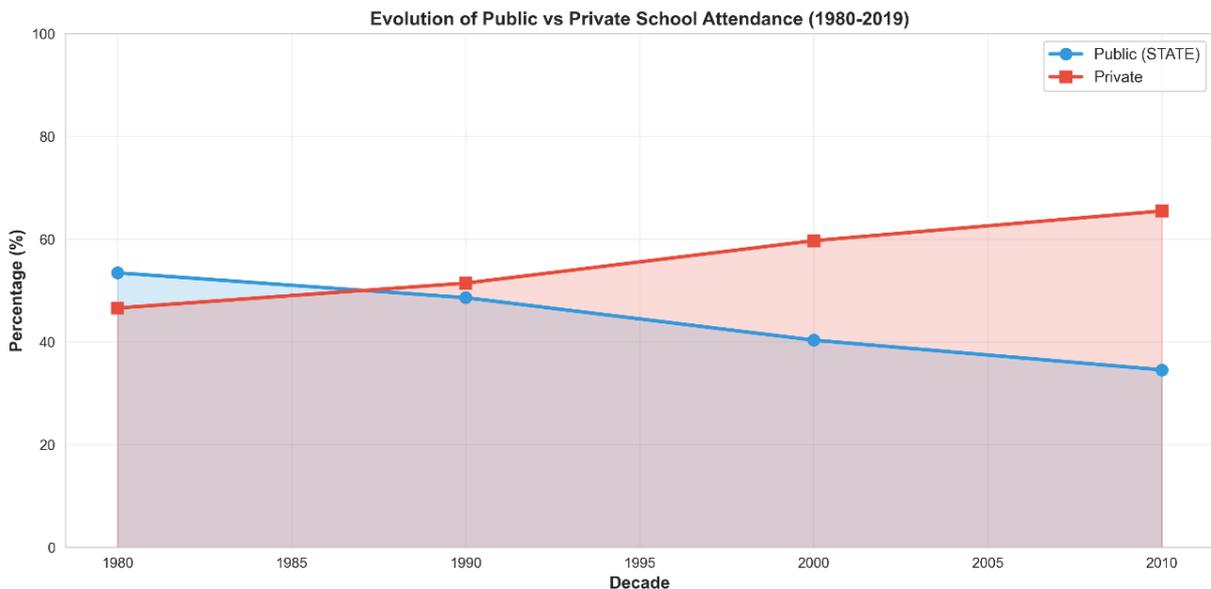

**Table 3. Composition of entering cohorts by school management (Public vs. Private) and decade.**

| Decade | Private | Public |
|---|---|---|
| 1980 | 46.58 | 53.42 |
| 1990 | 51.42 | 48.58 |
| 2000 | 59.69 | 40.31 |
| 2010 | 65.47 | 34.53 |

Interpreting **private management as a high-SES proxy** and **public management as a low-SES proxy**, this trajectory implies a progressive tilt towards entrants from relatively more advantaged educational backgrounds, within the subset of students for whom school data are available. The fact that this shift unfolds over multiple decades suggests structural changes in the composition of those who manage to convert secondary schooling into engineering-related university access.

### 4.3. Geographic origin of entrants (Analyses 3–4)

Figure 4 and Table 4 summarise the **province-level distribution** of entrants by decade, while Figure 5 tracks the evolution of the top provinces over time. Among records with valid province information, **the home province dominates throughout** but its relative share increases:

- 1980s: 67.6% of known-province entrants from the home province (Tucumán), with 10–11% each from Salta and Jujuy.

- 1990s: 68.7% from Tucumán; 14.6% from Salta; 8.2% from Jujuy.

- 2000s: 75.1% from Tucumán; 10.3% from Salta; 6.7% from Jujuy.

- 2010s: 82.4% from Tucumán; ~5–6% each from Salta and Jujuy; ~2% from Santiago del Estero and from Buenos Aires.

**Figure 4. Heatmap of geographic origin of entrants by province and decade.**

**Figure 5. Longitudinal trends of the top 10 provinces of origin: Localisation of the catchment area.**

**Table 4. Geographic distribution of entrants: Home province, regional neighbours, and other jurisdictions.**

| province_name_canonical | 1970 | 1980 | 1990 | 2000 | 2010 |
|---|---|---|---|---|---|
| Tucumán | 0.0 | 50.0 | 694.0 | 4604.0 | 5139.0 |
| nan | 1.0 | 3850.0 | 3606.0 | 2959.0 | 2.0 |
| Salta | 0.0 | 8.0 | 147.0 | 634.0 | 351.0 |
| Jujuy | 0.0 | 8.0 | 83.0 | 411.0 | 353.0 |
| Santiago del Estero | 0.0 | 3.0 | 24.0 | 147.0 | 121.0 |
| Buenos Aires | 0.0 | 0.0 | 20.0 | 121.0 | 121.0 |
| Catamarca | 0.0 | 3.0 | 15.0 | 104.0 | 88.0 |
| Córdoba | 0.0 | 1.0 | 9.0 | 33.0 | 21.0 |
| Santa Fe | 0.0 | 0.0 | 4.0 | 20.0 | 11.0 |
| Santa Cruz | 0.0 | 0.0 | 2.0 | 13.0 | 8.0 |
| Mendoza | 0.0 | 0.0 | 2.0 | 10.0 | 7.0 |
| Río Negro | 0.0 | 0.0 | 2.0 | 8.0 | 7.0 |
| Capital Federal | 0.0 | 0.0 | 3.0 | 9.0 | 4.0 |
| Chaco | 0.0 | 0.0 | 1.0 | 11.0 | 4.0 |

Two patterns stand out. First, **the faculty's catchment area becomes increasingly localised**, with the home province's share among known-province entrants rising by roughly 15 percentage points between the 1980s and 2010s. Second, **neighbouring north-western provinces** (Salta, Jujuy, Santiago del Estero, Catamarca) maintain a persistent presence, even as their relative shares decline somewhat in the 2010s. Entrants from more distant provinces, including the Buenos Aires metropolitan area, remain numerically small throughout, though they are academically and symbolically visible.

These trends must be interpreted in light of the strong improvements in province coding over time (Section 4.1). When restricted to decades with high coverage (2000s and 2010s), the data show a faculty that is increasingly anchored in its immediate region, with **stable but minority flows from the broader national territory**.

**4.4. Socio-economic proxies beyond school management (Analysis 4)**

To synthesise the joint effect of school management and orientation into a minimal set of socio-economic proxies, we classify entrants according to the public–private divide and interpret this divide in light of the local schooling ecology. In the faculty's catchment area, private secondary schools include a small set of religious and bilingual institutions, often located in high-income residential zones such as Yerba Buena and serving families with the resources to afford selective, fee-paying education. By contrast, public schools in the region enrol a much broader social spectrum and, in aggregate, represent predominantly middle- and lower-middle-income households.

On this basis, we define:

- upper-middle-class proxy entrants as those coming from private secondary schools (religious or secular), and
- non-upper-middle-class proxy entrants as those coming from public secondary schools.

This is deliberately simple and does not claim to capture the full stratification within either sector. However, it aligns with the concrete reality that, in this context, attendance at specific private schools and residence in high-income municipalities such as Yerba Buena function as strong markers of upper-middle-class status, whereas public schooling aggregates more heterogeneous but generally less advantaged trajectories.

At the decade level, this proxy naturally mirrors the public–private split described above ( Table 5 and Figure 6):

- 1980s: 46.6% upper-middle-class proxy, 53.4% non-upper-middle-class proxy.
- 1990s: 51.4% upper-middle-class proxy, 48.6% non-upper-middle-class proxy.
- 2000s: 59.7% upper-middle-class proxy, 40.3% non-upper-middle-class proxy.
- 2010s: 65.5% upper-middle-class proxy, 34.5% non-upper-middle-class proxy.

**Figure 6. Evolution of socio-economic proxy groups: Entrants from private (high-SES proxy) vs. public (low-SES proxy) schools.**

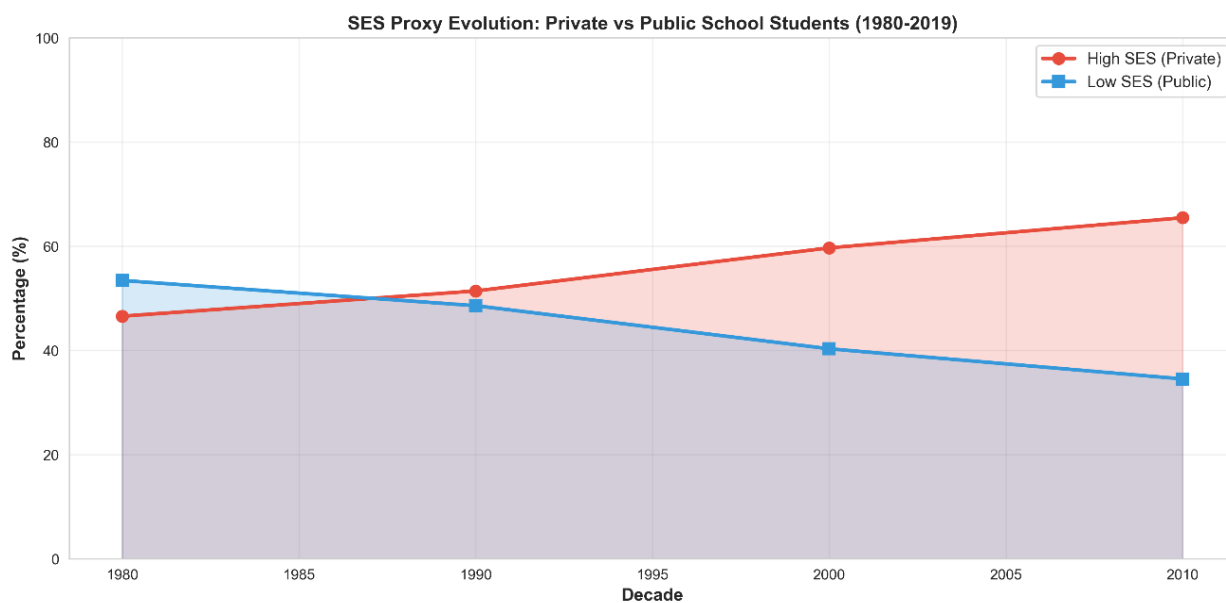

**Table 5. Evolution of socio-economic background proxies by decade.**

| Decade | High_SES | Low_SES |
|---|---|---|
| 1980 | 46.57 | 53.42 |
| 1990 | 51.41 | 48.58 |
| 2000 | 59.68 | 40.31 |
| 2010 | 65.46 | 34.53 |

While coarse, this proxy is robust to coding changes and available for the full documented window from the 1980s onwards. The monotonic increase in the upper-middle-class share suggests that, conditional on reaching the faculty, students from these more advantaged school and residential circuits have become progressively more prevalent relative to those from public schools. Later sections return to this pattern when discussing equity and the limits of gratuidad.

**4.5. Age at entry across decades (Analysis 5)**

Figure 7 and Table 6 describe the distribution of **age at first enrolment** by decade. Excluding the single 1970s case, the mean and median ages remain remarkably stable over time:

- 1980s: mean 19.7 years; median 19; IQR 18–20 (n = 3,926).
- 1990s: mean 20.4 years; median 19; IQR 19–20 (n = 4,643).
- 2000s: mean 20.1 years; median 19; IQR 19–20 (n = 9,212).
- 2010s: mean 19.8 years; median 19; IQR 18–20 (n = 6,341).

**Figure 7. Violin plots of age at entry distribution by decade, showing stability of central tendency and persistent right tails.**

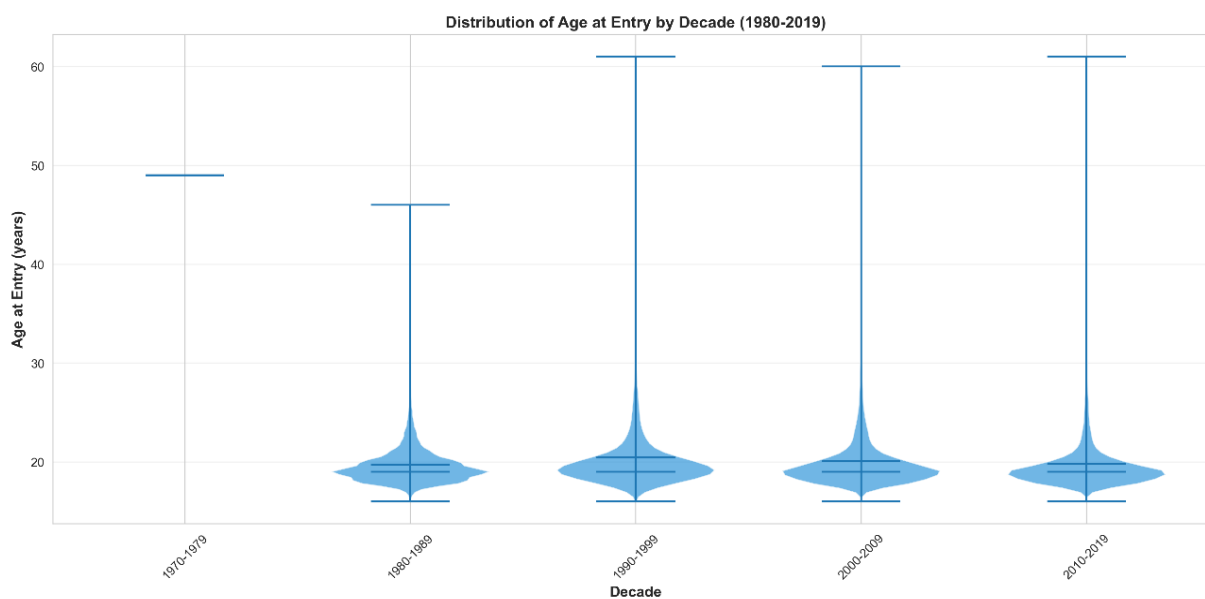

**Table 6. Descriptive statistics of age at first enrolment by decade (Mean, Median, IQR).**

| Decade | mean | median | std | p25 | p75 | count |
|---|---|---|---|---|---|---|
| 1970 | 49.0 | 49.0 | | 49 | 49 | 1 |
| 1980 | 19.72 | 19.0 | 2.19 | 18 | 20 | 3926 |
| 1990 | 20.43 | 19.0 | 3.76 | 19 | 20 | 4643 |
| 2000 | 20.08 | 19.0 | 3.2 | 19 | 20 | 9212 |
| 2010 | 19.79 | 19.0 | 3.08 | 18 | 20 | 6341 |

In all decades, the **median age is 19** and the interquartile range is narrow, indicating that the typical entrant arrives shortly after completing secondary school. However, the violin plots reveal **persistent long right tails** in every decade, with a non-trivial group of students enrolling in their mid-20s or later. The standard deviation hovers around 3–4 years, consistent with this mixture of "on-time" entrants and older students.

Taken together, these results suggest that, despite macroeconomic turbulence and policy changes, the **central tendency of age at entry has been stable**, while the institution has consistently served a minority of older entrants whose trajectories likely include delayed or interrupted schooling and labour-force participation.

## 4.6. Multivariate clustering of background profiles (Analysis 6: UMAP + DBSCAN)

To move beyond marginal distributions, we embed individual background profiles into a low-dimensional space using UMAP and then identify dense regions via DBSCAN (Figure 8, Figure 9). The clustering is performed on a subset of years and variables with high coverage, using one-hot encodings of entry decade, school management, school orientation, province group and main missingness patterns.

### 4.6.1. Cluster structure and noise

The UMAP + DBSCAN procedure identifies **126 clusters** plus one noise group, covering the **24,133 entrants** in N1c. The noise cluster (DBSCAN label –1) contains **309 students (≈1.3%)**, which lie in sparsely populated regions of the embedding. The remaining clusters range in size from a handful of students to several hundred.

**Figure 8. UMAP embedding of student background profiles coloured by DBSCAN cluster assignment.**

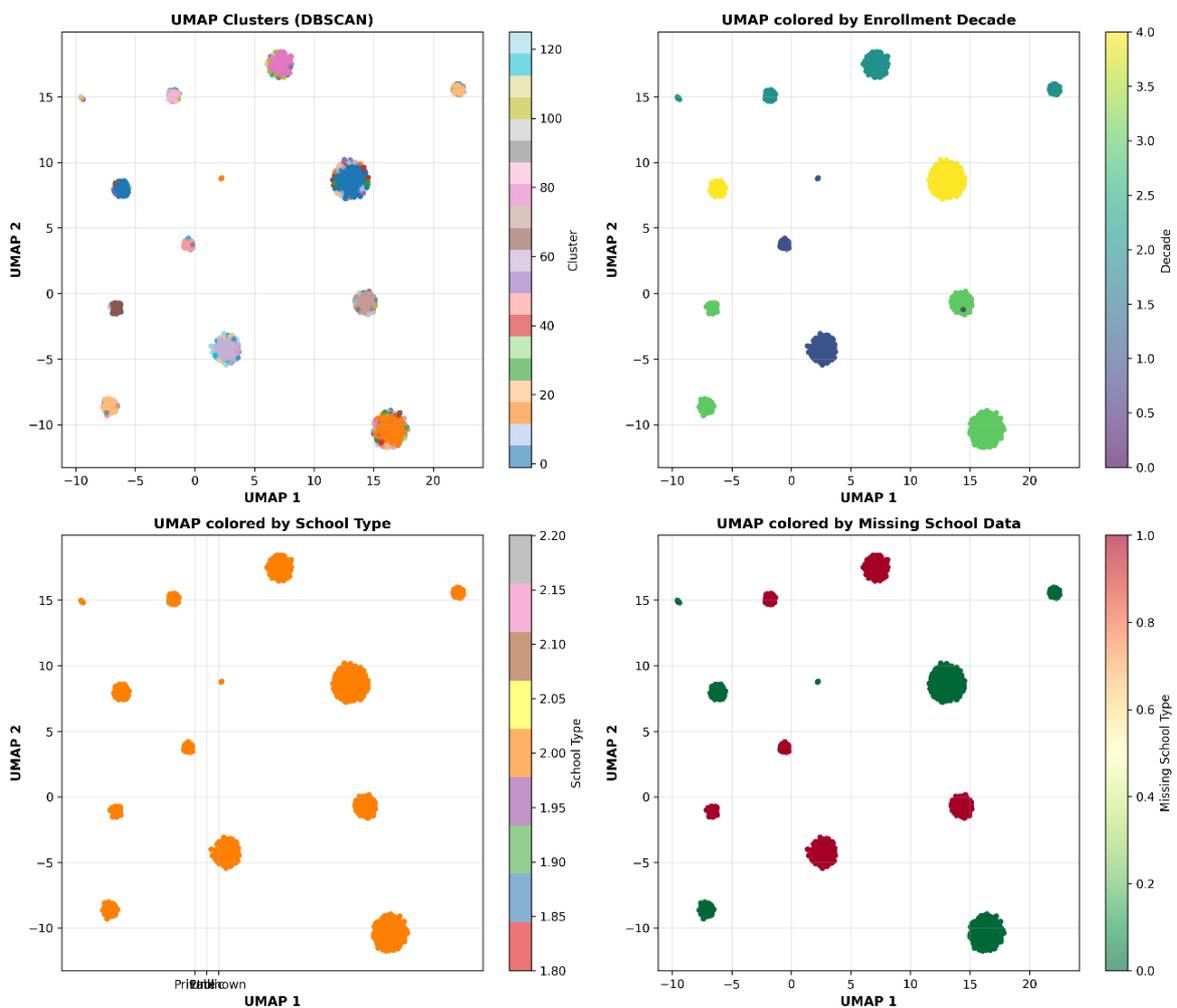

**Figure 9. UMAP embedding coloured by entry decade, revealing distinct temporal regimes in cohort composition.**

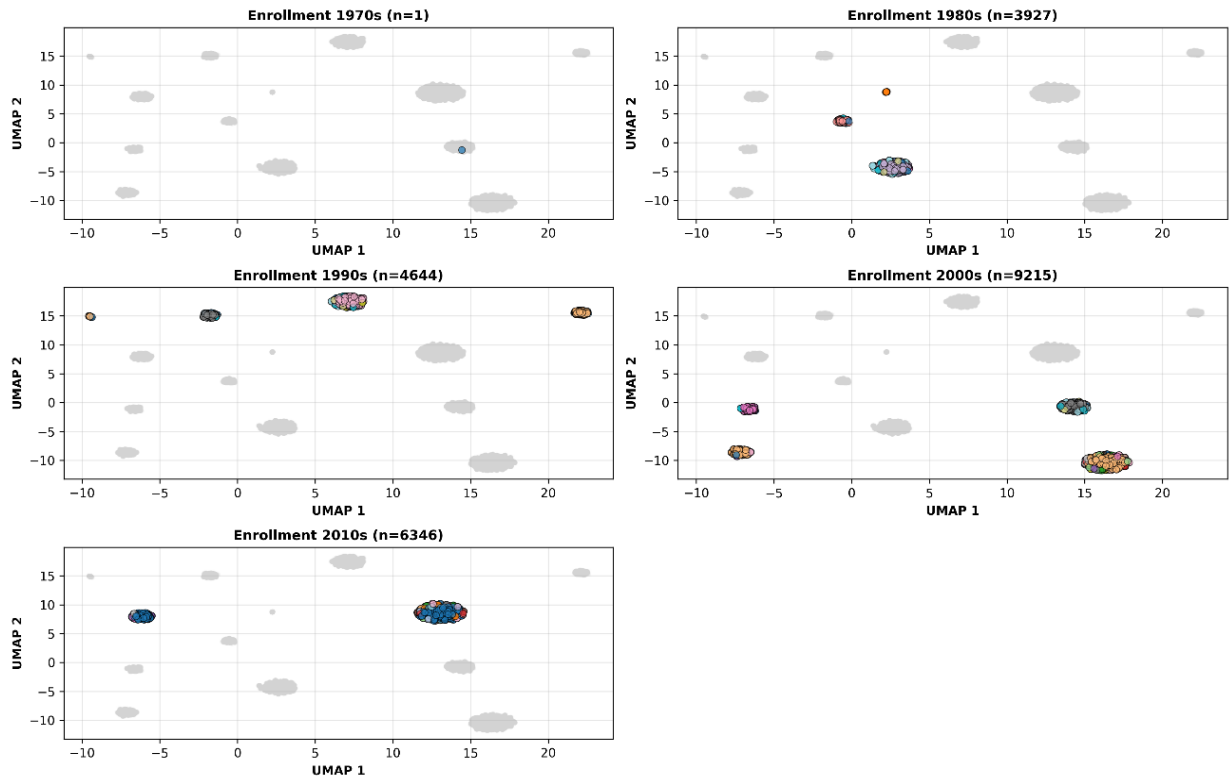

### 4.6.2. Decade-based regimes

When the UMAP embedding is coloured by **entry decade**, clusters align strongly with temporal regimes rather than forming a single, time-invariant cloud of "engineering students". In the 1980s and early 1990s, clusters are dominated by profiles with:

- high missingness in school variables,
- weaker geographic coding, and
- a mix of programme families.

From the 2000s onward, clusters become more sharply defined along **school management and orientation** (public/private, technical/non-technical) and **province group**, with almost no missingness flags. Figure 9 shows that each decade occupies partly distinct regions in the UMAP space, even after controlling for the types of variables used.

### 4.6.3. Cluster composition

Cross-tabulating cluster membership by decade, school sector and province group reveals several interpretable configurations, including:

- clusters concentrated in the **home province + private religious schools** in the 2000s–2010s,
- clusters mixing **interior provinces + public technical schools** in intermediate decades, and
- residual clusters associated with rare combinations of background attributes.

The key message for this manuscript is not the taxonomy of all 126 clusters, but the **clear evidence that background configurations are path-dependent and decade-specific**, reinforcing the idea that cohort composition must be understood in its historical trajectory rather than as a timeless snapshot.

### 4.7. Macroeconomic context and access patterns (Analysis 7: Macro–N1c linkage)

The final analysis overlays the N1c cohort summaries with the reconstructed macroeconomic panel for Argentina (1980–2019). Figure 10 presents a multi-panel view of:

- inflation and unemployment trends,
- the share of entrants from interior provinces,
- the share from private schools, and
- average age at entry, all aligned by year of entry.

**Figure 10. Multi-panel association view: Macroeconomic indicators (inflation, unemployment) vs. cohort characteristics (private school share, interior origin, age at entry).**

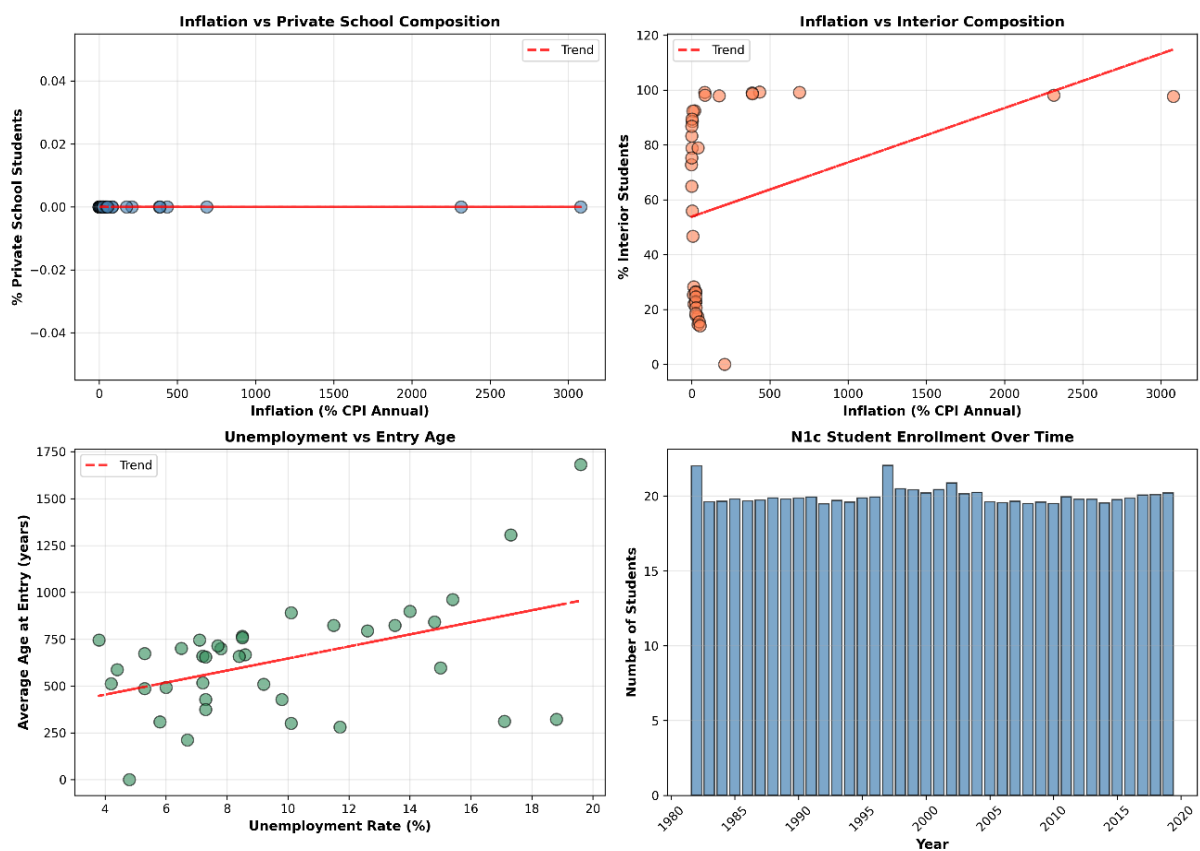

**Table 7. Bivariate correlations between national macroeconomic indicators and cohort composition variables (1980–2019).**

| macro_variable | n1c_variable | n_obs | pearson_r | pearson_p | spearman_r | spearman_p |
|---|---|---|---|---|---|---|
| inflation_cpi_annual_pct | pct_interior | 38 | 0.345 | 0.033 | 0.192 | 0.248 |
| unemployment_rate_pct | avg_age | 38 | 0.457 | 0.003 | 0.374 | 0.0203 |

At the descriptive level, Argentina's macro series exhibit the expected features: episodes of very high inflation (including hyperinflation peaks above 3,000% in the late 1980s), subsequent disinflation under the convertibility regime, the 2001–2002 crisis, and renewed volatility in the 2010s. Unemployment and poverty also show marked cyclical movements, particularly around major crises.

When correlated with N1c cohort characteristics (Table 7), two associations stand out:

- **Inflation and geographic composition.** The annual inflation rate at entry is **positively correlated** with the **percentage of entrants from interior provinces** (Pearson $r = 0.35$, *p* = 0.034; Spearman $\rho = 0.19$, *p* = 0.25; *n* = 38). Periods of higher inflation tend to coincide with somewhat larger shares of students coming from outside the home province, although rank-based association is weaker.

- **Unemployment and age at entry.** The unemployment rate at entry year is **moderately and significantly correlated** with **average age at entry** (Pearson $r = 0.46$, *p* = 0.0039; Spearman $\rho = 0.37$, *p* = 0.020; *n* = 38). Years with higher unemployment tend to see slightly older entering cohorts, consistent with the idea that labour-market conditions and schooling decisions are intertwined.

By contrast, the **share of private-school entrants** shows no strong or consistent correlation with either inflation or unemployment in this descriptive framework, suggesting that the long-run drift towards private-school backgrounds is driven more by structural changes in the secondary schooling system and social stratification than by year-to-year macroeconomic shocks.

These results do not support causal claims, but they provide **quantitative anchors** for the narrative that follows: access to engineering and science programmes in this faculty has evolved under the joint influence of changing school systems, institutional practices and a macroeconomy characterised by recurrent turbulence, with detectable but selective links between macro indicators and who manages to enrol.

## 5. DISCUSSION

### 5.1. Free tuition, open access and persistent inequality

At first sight, the faculty analysed here operates under conditions that should favour equity: **no tuition fees, no numerus clausus at entry and an explicit public mission** aligned with the Argentine principle of free higher education. Yet the reconstructed background layer shows that, over four decades, the composition of entering cohorts has shifted towards students from **upper-middle-class school and residential circuits**, with only modest evidence of diversification towards more vulnerable profiles.

Within the subset of students for whom school management is observed, the share of entrants from private secondary schools rises from less than half in the 1980s to roughly two-thirds in the 2010s. In the concrete geography of the faculty, this does not simply mean "private" in the abstract. It refers, to a significant extent, to

**religious and bilingual colegios located in high-income areas such as Yerba Buena**, whose fee structures, language offerings and peer composition are characteristic of upper-middle-class households. Public schools, by contrast, draw predominantly from middle- and lower-middle-income families and from more heterogeneous neighbourhoods.

At the same time, the faculty's catchment area becomes more local, with a growing concentration of entrants from the home province and a declining share from other provinces. Taken together, these trends indicate that **free tuition and open doors at the university level coexist with strong upstream filters** in the school system and in territorial opportunity structures. Students who arrive at the faculty are increasingly those who have passed through high-resource private institutions and reside in relatively affluent municipalities, rather than those from the most disadvantaged sectors that gratuidad is often presumed to favour.

This pattern resonates with national and regional analyses showing that **unrestricted access and the absence of fees do not, by themselves, produce a socially representative student body**. In Argentina and across Latin America, access to high-return public universities remains skewed towards students with stronger school trajectories, greater cultural capital and more economic security, even under formally inclusive admission regimes. The case examined here therefore illustrates a broader point: **gratuidad is a necessary baseline for equity, but it is far from sufficient** when the pipelines feeding public universities are already stratified by class and territory.

### 5.2. Massification, stratified pipelines and local equity regimes

From a system perspective, Argentina fits within the global trend towards **high participation systems** of higher education, where a large share of each cohort enters some form of tertiary education (Marginson, 2016; Cantwell et al., 2018). Latin-American analyses highlight that this expansion has been accompanied by **growing institutional and social stratification**, including the rapid growth of private provision and internal differentiation within the public sector (Ferreyra et al., 2017; Labraña & Brunner, 2022).

The faculty studied here is not at the apex of the national prestige hierarchy, but it is located in a **public research university** and offers engineering and science programmes with relatively high symbolic and labour-market value. In this context, our findings suggest the existence of a **stratified pipeline**:

- At the **secondary level**, students from private (especially religious) schools become increasingly dominant among those who eventually enter the faculty, even as public education remains the main supplier of upper-secondary graduates in the province and the region.

- At the **territorial level,** the system appears to serve primarily the local middle classes and selected groups from nearby provinces, rather than functioning as a broad regional or national equaliser.

This is consistent with empirical work showing that, despite expansion and free tuition, **access to high-return public universities in Argentina remains skewed towards students with higher socio-economic status and better school trajectories** (Adrogué & García de Fanelli, 2021; García de Fanelli, 2010). In Trow's terms, the system may have moved from **elite to mass participation**, but without resolving the underlying **"sponsored mobility"** logics that channel more advantaged students into the most valued institutions and programmes.

At the faculty scale, the combination of free tuition with increasingly stratified background profiles implies that **equity must be understood as a local regime**, not an automatic property of the national policy framework. The same national rule of free and open public universities can produce different degrees of social selectivity depending on local school ecologies, information flows, and the internal culture of programmes.

### 5.3. Macro shocks, age at entry and the limits of "gratuidad"

The macro–N1c linkage adds a temporal layer to this picture. Two patterns are particularly suggestive.

First, **years with higher unemployment are associated with slightly older entering cohorts**, with statistically significant correlations in both Pearson and Spearman metrics. This suggests that in adverse labour markets some young adults may **re-enter or postpone higher education**, using university studies as a shelter or as a way to re-position themselves, while others may be pushed away from study and into work. Free tuition lowers the direct monetary cost of such decisions, but it does not equalise the opportunity cost of studying for students who must contribute to household income.

Second, **periods of higher inflation show a positive correlation with the share of entrants from interior provinces**, although this association is weaker. One interpretation is that macroeconomic turbulence alters **migration and educational strategies** within the region, perhaps encouraging families in nearby provinces to send their children to perceived "safer" public institutions when private options become less affordable in real terms. Another is that inflation interacts with the timing of public policies (e.g., scholarship schemes, transport subsidies), which may temporarily widen or narrow the effective catchment radius of the faculty.

These findings echo regional work arguing that **macro-structural conditions and student finance regimes are jointly responsible for equity outcomes**, even in formally free systems (García de Fanelli, 2019; Ferreyra et al., 2017). In a context of recurrent crises, the promise of gratuidad is constantly tested by **household-level liquidity constraints, unstable labour markets and the need for students to work**, which may limit the benefits of tuition-free policies for those in the lowest income deciles.

**5.4. Policy and institutional implications**

Taken together, the results support a clear, if uncomfortable, conclusion: **in this faculty, tuition-free and open access policies have not been sufficient to guarantee equity in who enters engineering and science programmes**. Over four decades, the observed trend is towards:

- greater representation of students from private secondary schools,
- stronger localisation of the catchment area, and
- persistent age stratification with a stable on-time majority and a smaller, older group whose trajectories likely intersect with labour-market volatility.

These patterns do not negate the value of free public universities; rather, they show that **equity requires more than the absence of tuition fees and formal restrictions**. Three implications follow.

1. **Equity must be monitored at the intra-institutional level.** Relying on national averages or on the normative appeal of gratuidad obscures local stratification patterns. Leakage-aware, cohort-based analytics such as those used here can help faculties track who is entering over time, with enough granularity to inform targeted outreach and support.

2. **Upstream interventions matter.** If the pipeline is increasingly dominated by private-school graduates, policies that rely solely on university-level financial aid or tutoring will have limited redistributive impact. The literature on Latin America suggests that **reducing inequality in access to quality secondary education** and strengthening transitions for low-SES students are indispensable complements to tuition-free university systems.

3. **Macro-sensitive support mechanisms are needed.** The association between unemployment and older age at entry, and between inflation and changes in geographic composition, indicate that **cohort profiles respond to macro shocks**. Faculties that wish to honour the promise of equal opportunity under gratuidad may need **counter-cyclical instruments** (emergency scholarships, flexible study–work arrangements, targeted support for non-traditional entrants) that explicitly recognise these dynamics.

For the broader CAPIRE programme, these findings underscore the importance of **embedding equity questions into the core of data architecture and modelling**, rather than treating them as ex post add-ons. The same leakage-aware background layer that feeds predictive models of dropout can also serve as a **mirror** in which institutions confront the social selectivity of their formally open gates. In that mirror, tuition-free access appears as a necessary baseline—but very far from a sufficient condition—for the kind of equity that mass higher education systems now claim to pursue.

## 6. LIMITATIONS AND ROBUSTNESS

### 6.1. Constraints of administrative data

The analyses presented in this article rest on an administrative dataset that was **not originally designed for research purposes**. As in many universities, background variables such as secondary school type, school management and province of origin were captured for operational reasons, through evolving forms and legacy systems, rather than as part of a coherent longitudinal research design. This has three implications.

First, **coverage and precision vary across decades**. The earliest years of the series show very high missingness in key variables, followed by gradual improvements as information systems were standardised. Although our missingness analysis indicates that gaps are structurally patterned rather than random, this does not eliminate the fact that early decades are **less richly described** than later ones. Second, some potentially relevant dimensions—such as parental education, household income, or detailed work–study arrangements—are absent from the administrative records and therefore cannot be reconstructed. Third, coding choices at the time of data entry (e.g., how schools were labelled, how provinces were abbreviated) may have introduced subtle biases that even careful normalisation cannot fully remove.

These limitations are not unique to this case; they reflect broader concerns in the use of administrative data for educational research, where trade-offs between scale, detail and data quality are pervasive (Card et al., 2010; Huebener & Marcus, 2020). Enrolment records offer full coverage of students who reach the institution, but they observe only a fragment of their earlier trajectories.

### 6.2. Structural missingness and conservative analytic choices

A central feature of the dataset is the **structural missingness** in school-type and geographic variables for early decades. Our logistic models show that missingness is highly predictable from entry decade, degree family and basic coding features, which supports treating it as historically structured rather than as MCAR noise. However, this does not imply that we can **recover** the missing information; it simply means that we can **characterise** its pattern.

To minimise the risk of introducing artefacts, we adopt **conservative decisions**:

- No multiple imputation is performed for the core descriptive results.
- All percentages explicitly report the number of cases with observed data in the denominator.
- Early decades with extremely low coverage are interpreted cautiously, and we avoid fine-grained subgroup analyses where denominators become unstable.

These choices increase transparency at the expense of statistical efficiency. A different modelling strategy—e.g., using multiple imputation under MAR assumptions (Little & Rubin, 2019) or partial pooling across decades in a hierarchical framework (Gelman & Hill, 2006)—could in principle extract more information from the same data, but would also embed stronger assumptions. For this first background-focused manuscript in the CAPIRE 2 series, our priority is to provide **robust descriptive baselines** with minimal modelling overhead. More elaborate treatments of missingness will be more appropriate in downstream work on predictive or causal modelling.

**6.3. Internal validity and measurement error**

Within each decade, measurement error may arise from:

- **Misclassified school types or management** (e.g., schools that changed administrative status over time but were recorded under outdated labels).
- **Ambiguous or inconsistent province coding**, particularly for students who moved during their schooling.
- **Age at entry** computed from administrative birthdates that may contain occasional errors.

Our harmonisation pipeline—combining dictionary-based cleaning, fuzzy matching for school names and aggregation into broader categories—is designed to **stabilise** classifications over time rather than to recover every fine-grained distinction. This means that some internal variation (for instance, among different kinds of private schools) is deliberately collapsed into robust high-level categories. From a measurement perspective, we privilege **coarse but reliable proxies** over fine but noisy distinctions, in line with recommendations for working with historical administrative data (Huebener & Marcus, 2020).

These decisions likely attenuate some associations. If misclassification is roughly symmetric, it will tend to **bias estimated relationships towards zero** rather than creating spurious patterns. The strong and monotonic trends observed in private versus public shares, and in the localisation of the catchment area, therefore represent **lower bounds** on the underlying shifts rather than artefacts of the cleaning process.

**6.4. External validity and institutional specificity**

The data used here come from a **single faculty of engineering and exact sciences** in a public university in north-western Argentina. As such, the results are not intended to represent all Argentine higher education, nor even all engineering programmes in the country. Institutional histories, local labour markets and regional school ecologies differ substantially across universities (Ferreyra et al., 2017). The specific configuration of free tuition, open access, programme mix, and territorial location analysed here constitutes one **local equity regime** among many.

Nevertheless, the patterns we identify—growing dominance of private-school backgrounds among entrants, increasing localisation of the catchment area, and selective links to macroeconomic indicators—echo findings from national-level and comparative work on access and stratification in Latin American higher education (Arias Ortiz et al., 2020; García de Fanelli, 2019; Marginson, 2016). This suggests that, while the exact magnitudes may be institution-specific, the **direction of change** and the core tensions around equity under gratuidad have broader relevance.

An additional external-validity limitation is that our macro–N1c linkage uses **national-level macro indicators**, which may not fully capture subnational economic conditions affecting the faculty's hinterland. Future work could refine this dimension by incorporating regional labour-market measures, local unemployment series or sectoral shocks.

**6.5. Descriptive focus and the scope of inference**

Finally, the analyses in this article are explicitly **descriptive**. We document covariations between background composition and macroeconomic context, but we do not estimate causal effects of inflation, unemployment or policy changes on who enters the faculty. Correlations between macro indicators and cohort characteristics may reflect a mixture of underlying mechanisms, including:

- direct effects on household decisions,
- concurrent policy interventions (scholarships, transport subsidies), and
- structural changes in the schooling system that coincide with macro cycles.

Causal inference in this setting would require **stronger research designs**, such as exploiting quasi-experimental shocks, using difference-in-differences or synthetic control methods, or embedding the macro–micro linkage in a formal structural model (Angrist & Pischke, 2009; Imbens & Rubin, 2015). Those approaches lie outside the scope of the present manuscript, whose goal is to **establish a baseline**: a coherent, leakage-aware description of who enters the faculty over four decades, and how these patterns sit alongside the country's turbulent macroeconomic history.

Within this descriptive scope, robustness comes from consistency across multiple lenses: simple decade-level summaries, multivariate clustering, and macro-linked correlations all point in the same qualitative direction. The limitations discussed above should temper over-interpretation, but they do not undermine the central conclusion: **in this faculty, free and open access has operated within, rather than against, long-standing structures of social and territorial inequality.**

## 7. CONCLUSIONS

### 7.1. Main findings

This article used a leakage-aware background layer (N1c) to reconstruct **four decades of entering cohorts** in a faculty of engineering and exact sciences in north-western Argentina, linking student backgrounds to a documented macroeconomic panel. Several conclusions emerge.

First, the analysis of **structural missingness** shows that early-decade gaps in school-type and geographic data are historically patterned, not random. Missingness is almost universal in the 1980s, declines sharply in the 1990s and 2000s, and becomes negligible in the 2010s. This justifies treating the early period as a partially observed regime, and motivates the conservative, decade-based summaries used throughout the paper.

Second, within the documented regime, **school-sector composition has shifted steadily towards private secondary schools**, especially private religious institutions. Among students for whom school management is known, the share from private schools rises from less than half in the 1980s to roughly two-thirds in the 2010s, implying a growing high-SES tilt among entrants to engineering and science programmes.

Third, the faculty's **catchment area has become more local**. The home province gains weight among entrants with known province, while neighbouring provinces remain present but decline proportionally, and distant provinces play only a minor role. Free and open access is thus operating within a primarily regional ecosystem, with limited evidence of broad territorial diversification.

Fourth, **age at entry is remarkably stable**: median age remains at 19 in all decades, with narrow interquartile ranges. At the same time, long right tails in each decade reveal a persistent minority of older entrants, whose trajectories likely involve delayed or interrupted schooling and interactions with the labour market.

Fifth, the **UMAP + DBSCAN analysis** shows that background configurations are not time-invariant. Clusters are strongly structured by entry decade, with early cohorts characterised by high missingness and weaker geographic coding, and later cohorts by sharper splits along school sector and province group. There is no single, timeless profile of "the engineering student"; instead, there are **decade-specific regimes** that reflect shifts in schooling systems and institutional practices.

Finally, the **macro–N1c linkage** reveals selective associations rather than a tight coupling between context and access. Higher unemployment at entry is associated with slightly older cohorts, and higher inflation correlates with a larger share of students from interior provinces, while the long-run drift towards private-school entrants appears largely decoupled from year-to-year macroeconomic swings. These patterns are descriptive, but they underscore that **free tuition and open access have evolved under recurrent macroeconomic turbulence, without erasing underlying social and territorial stratification**.

**7.2. Implications for future research**

The findings point to several directions for future work within the CAPIRE 2 programme.

First, the N1c background layer should be **linked explicitly to trajectory outcomes**—dropout, migration between programmes, time to degree and graduation—to quantify how background regimes translate into differential chances of success across engineering disciplines. This will allow the project to move from "who gets in" to "who gets through", under the same leakage-aware principles.

Second, the **cross-programme architecture** of CAPIRE 2 opens the door to comparing background patterns and outcomes across multiple engineering degrees, rather than within a single faculty aggregate. Differences in curriculum topology, assessment cultures and institutional selectivity may interact with the background shifts documented here, producing distinct equity profiles by discipline.

Third, the macro panel and the structural shocks already identified in CAPIRE (e.g., teacher strikes, major crises) can be embedded in **causal and simulation models** that go beyond descriptive correlations. Quasi-experimental designs and agent-based models will be needed to test how specific policies—such as targeted scholarships, articulation agreements with public technical schools or changes in admission procedures—might alter the composition and trajectories of future cohorts under realistic macroeconomic scenarios.

Finally, the leakage-aware pipeline and the indicators developed here can be generalised into a **routine equity-monitoring toolkit** for faculties and universities: a small set of background metrics (school sector, territorial origin, age and SES proxies) tracked by cohort and decade, anchored in transparent data engineering rather than ad hoc reports.

### 7.3. Closing remarks

Taken together, the results portray a faculty that has honoured the **principle of gratuidad** in formal terms—no tuition fees, open entry, and a public mission—but within a **stratified pipeline** that increasingly favours students from **upper-middle-class school and residential circuits**. Over time, entrants are more likely to come from fee-paying religious and bilingual colegios located in high-income areas such as Yerba Buena, and less likely to originate from the more heterogeneous and generally less advantaged public-school sector or from distant provinces. Free and open access has been a necessary baseline; it has not been sufficient to ensure that those who arrive at the doors of engineering and science programmes reflect the full social and territorial diversity of the region.

By turning decades of "administrative chaos" into **analytical cohorts**, this manuscript provides a baseline from which more ambitious questions can be asked. Subsequent CAPIRE 2 studies will build on this layer to model trajectories, archetypes, shocks and interventions, and to examine how these upper-middle-class entry patterns relate to differential risks of dropout and graduation across programmes. The starting point, however, is simple and non-negotiable: before predicting who will drop out or simulating policy interventions, we need to understand **who was effectively able to enter the game**—and how that answer has shifted over forty years of massified yet unequal higher education in Argentina.

## REFERENCES

Adelman, M., Haimovich, F., Ham, A., & Vázquez, E. (2018). Predicting school dropout with administrative data: New evidence from Guatemala and Honduras. Education Economics, 26(4), 356–372. https://doi.org/10.1080/09645292.2018.1433127